\documentstyle[pra,aps]{revtex}
\begin{document}
\draft
\title{Momentum Distribution in Nuclear Matter and Finite Nuclei}
\author{H. M\"uther, G. Knehr}
\address{
Institut f\"ur Theoretische Physik, Universit\"at T\"ubingen,\\
Auf der Morgenstelle 14, D-72076 T\"ubingen, Germany  }
\author{A. Polls}
\address{Departament d'Estructura i Constituents de la Mat\`eria\\
Universitat de Barcelona\\
Diagonal 647, E-08028 Barcelona, Spain}
\date{\today}
\maketitle
\begin{abstract}
A simple method is presented to evaluate the effects of short-range
correlations on the momentum distribution of nucleons in nuclear
matter within the framework of the Green's function approach. The method
provides a very efficient representation of the single-particle Green's
function for a correlated system. The
reliability of this method is established by comparing its results to
those obtained in more elaborate calculations. The sensitivity of the
momentum distribution on the nucleon-nucleon interaction and the
nuclear density is studied. The momentum distributions of nucleons in
finite nuclei are derived from those in nuclear matter using a local-density
approximation. These results are compared to those obtained directly
for light nuclei like $^{16}O$.
\end{abstract}
\pacs{PACS numbers: 21.10.Jx, 21.10.Pc, 21.65.+f, {\bf 24.10Cn}}

\section{Introduction}
Realistic nucleon-nucleon (NN) interactions like the phenomenological
Reid soft-core potential\cite{reid} or One-Boson-Exchange (OBE)
potentials\cite{rupr}, which are adjusted to fit the NN scattering data,
typically contain rather strong short-range components. These
short-range parts as well as a non-negligible tensor component are
responsible for the fact that simple mean-field or Hartree-Fock (HF)
calculations of nuclear systems yield very unsatisfactory results.
It turns out that HF calculations using such realistic NN forces may not
even lead to bound nuclei\cite{prog2}. Therefore, based on these
theoretical considerations it seems obvious that nuclear wave-functions
must contain correlations, which are induced by these short-range and
tensor components and cannot be accounted for in the mean-field or HF
approximation to the solution of the many-body problem.

The question is, whether there exist experimental observables which
reflect these correlations in an unambiguous way. In particular it would
be nice if one could explore these correlations in terms of single-nucleon
observables since they are easier to measure as well as to calculate.
This leads to the question how correlations affect the single-particle
density $\rho(r,r')$ in the nuclear many-body system. Rather
than discussing this non-local representation of the density matrix, we
may consider as well its Wigner transform\cite{ring} $f(R,k)$. Integrating
this Wigner distribution over all momenta $k$ yields the local density
$\rho(r=r')$. This local density distribution, or to be more precise
the corresponding charge distribution has been investigated with high
precision in elastic electron scattering experiments\cite{fried}. Also
the matter distribution can be analyzed in a rather model-independent
way by means of elastic $\alpha$ scattering and other
probes\cite{gils}. It seems, however, that these "experimental" charge
and matter distributions can very well be reproduced within a
mean-field approximation for the nuclear wave-function.

Integrating the Wigner transform of the one-nucleon density matrix over
all spatial coordinates one obtains the momentum distribution $n(k)$.
For an infinite system, invariant under local transformations, the
mean-field or Hartree-Fock prediction for this momentum distribution is
identical to the momentum distribution of a free Fermi gas. This means
that all states with momenta less than the Fermi momentum $k_{F}$ are
occupied with a probability $n=1$, while all states with momenta $k$
above $k_{F}$ are completely unoccupied ($n=0$). Correlations beyond
the HF approach modify this momentum distribution in the sense that
states with momenta below $k_{F}$ are partly depleted, whereas states
with high momenta are partly occupied.

{}From these considerations for the infinite system of nuclear matter one
may expect that correlations beyond HF will enhance the momentum
distribution at high momenta $k$ also for finite nuclei. This is one
reason why modern electron accelerators have been used to explore the
momentum distribution of nucleons in nuclei by means of nucleon
knock-out, $(e,e')p$, experiments\cite{lapi,bobel,blom}.

Microscopic nuclear structure calculations which account for the
effects of short-range and tensor correlations of realistic NN
interactions are mainly performed for very light
nuclei\cite{morita,tadok,ciofi,benpa} or infinite nuclear
matter\cite{mahau,ramos,benff,ciofb,vonde,gearh,baldo,kohl}. From these
results for nuclear matter one then tries to extract the effects of NN
correlations in order to estimate their influence on the momentum
distribution of real nuclei using a local density approximation
(LDA)\cite{stringari,neck,benh}.

Recently, there have also been attempts to determine the momentum
distribution in a microscopic calculation considering directly finite
nuclei as $^{16}O$\cite{pieper,pap1,pap2}. It turns out that the Green's
function approach is particularly useful for these investigations.
This method not only provides the total momentum distribution but also
yields detailed information on the spectral function. This spectral
function contains the information at which excitation of the residual
nucleus or, using the nomenclature of the knock-out experiment, at which
missing energy the various components of the momentum distribution
should be observed. These studies predict that high-momentum components
in the momentum distribution due to short-range correlations should
show up preferentially at large missing energies.

The comparison of the momentum distributions obtained within the
Green's function approach
for $^{16}O$ with corresponding ones for nuclear matter exhibited
remarkable discrepancies\cite{pap2}. Therefore the question arises if these
discrepancies demonstrate the limitation of the LDA in predicting the
momentum distribution. As it has been argued already in \cite{pap2}, such
a conclusion would be premature since that comparison was plagued with
various inconsistencies like e.g.
\begin{itemize}
\item The momentum distribution calculated for nuclear matter using the
Green's function approach has been
available only for one specific density, the saturation density of
nuclear matter. This density may be too large to be typical for the
situation of nuclei as small as $^{16}O$.
\item The result for nuclear matter has been derived from a
self-consistent calculation of the single-particle Green's
function\cite{vonde} whereas the calculation for the finite system has
been performed considering contributions to the self-energy of the
nucleons up to second order in a nuclear matter G-matrix\cite{pap2}
\item The calculation of the self-energy for the nucleons in the finite
system has been made employing a single-particle spectrum with a
substantial gap at the Fermi surface, whereas a continuous prescription
has been used for nuclear matter.
\item The calculation for the finite system was limited to partial
waves with angular momenta $l\leq 3$. This limitation may be too severe
for the momentum distribution at high momenta.
\item While the Reid soft-core potential\cite{reid} has been used for the
study of nuclear matter in \cite{vonde} the OBE potential $B$ of
\cite{rupr} has been employed for the calculation of $^{16}O$.
\end{itemize}

It is one aim of the present investigation to remove some of the
differences between these calculations of finite nuclei and infinite
nuclear matter. Furthermore we want to study the sensitivity of the
calculated momentum distributions in the nuclear systems on the nuclear
density, the NN interaction considered and various other ingredients of
the many-body calculation. For that purpose we have developed a new
very efficient method to calculate the momentum distribution in nuclear
matter
using an approximation to the Green's function method very similar to
the one presented in \cite{pap2} for finite nuclei. The comparison of
results obtained with this approximation scheme with those resulting
from the much more sophisticated scheme of \cite{vonde} demonstrates
the reliability of the method developed here. The efficiency of the new
scheme allows the detailed studies mentioned above.

After this introduction section 2 of this paper describes the
technique to be used for studies of nuclear matter. In particular we
will also present an efficient representation of the single-particle
Green's function, which allows a self-consistent treatment. In section 3
we briefly review the basic approximations used in the calculation of
the momentum distribution for finite nuclei of \cite{pap2} and
we outline a method to determine this momentum distribution,
in which the mean field part is calculated for the finite
system but the effects of correlations are taken from nuclear matter at
various densities, using a LDA. The results of the numerical calculations
are presented in section 4. In this chapter we discuss the
sensitivity of the results in nuclear matter on the various
ingredients. Also we compare the predictions of the LDA with
results obtained by the method of \cite{pap2}. For that purpose we
extended the studies of \cite{pap2} by considering different
interactions and allowing for higher partial waves. The main
conclusions are summarized in section 5.

\section{Momentum Distribution in Nuclear Matter}

\subsection{Self-Energy and Dyson Equation}
Our calculation of the single-particle Green's function for nucleons in
nuclear matter is based on the definition of the self-energy of the
nucleon, which includes the terms of first and second order in an
effective interaction ${\cal V}$, which we will define below. The
expression for the term of first order, displayed in
Fig.\ref{fig:diag}a, corresponds to the Hartree-Fock expression for the
single-particle energy of a nucleon with momentum $k$ in a system of
nuclear matter with a Fermi momentum $k_{F}$
\begin{eqnarray}
\Sigma^{(HF)} (k) & = &\sum\!\!\!\!\int_{h<F} <kh \vert {\cal V} \vert kh >
\nonumber \\
& = &\sum_{LSJT} (2J+1) (2T+1) \biggl\lbrace \Theta (k_{F}-k)
\int_{0}^{\frac{1}{2} \vert k-k_{F}\vert} dq\, 8q^2 \nonumber \\
&&+ \frac{1}{k} \int_{\frac{1}{2} \vert k-k_{F}\vert}^{\frac{1}{2}
( k+k_{F} )} dq \, q\left[ \left( k_{F}^2-q^2 \right) -4q\left( q-k_{F}\right)
\right] \biggr\rbrace {\cal V}_{LL}^{JST} (K_{av};q,q).
\label{eq:selfhf}
\end{eqnarray}
In the second part of this equation the matrix elements of ${\cal V}$
are given using the conventional partial wave representation with $L$,
$S$, $J$ and $T$ denoting the orbital angular momentum for the relative
motion, the spin, the total angular momentum and the isospin of the two
interacting nucleons, respectively. The relative momentum $q$ is
diagonal and an average value $K_{av}$ has been used for the c.m.
momentum, which is given by\cite{haftel}
\begin{equation}
K_{av}^2 = \cases{k^2+q^2, & if $0\leq q \leq {1\over 2}\vert
k-k_{F}\vert $ \cr {3 \over 4}k^2 -qk+{1\over 4} k_{F}^2, & if ${1\over 2}
\vert k-k_{F}\vert \leq q \leq {1\over 2} ( k+k_{F})$ \cr}
\end{equation}
The term of second order in the effective interaction with intermediate
2-particle 1-hole (2p1h) states, displayed in Fig.ref{fig:diag}b, also
depends on the energy $\omega$ of the nucleon under consideration and can
be calculated according to
\begin{eqnarray}
\Sigma^{(2p1h)}(k,\omega ) & = & \sum\!\!\!\!\!\!\int_{h<F}
\sum\!\!\!\!\!\!\int_{p_{1},p_{2}>F} \frac {< kh \vert {\cal V}\vert
p_{1},p_{2}>^2}{\omega - (\epsilon_{p1}+\epsilon_{p2}-\epsilon_{h}) +
i\eta }\nonumber \\
& = &\sum_{LSJT} (2J+1) (2T+1) \biggl\lbrace \Theta (k_{F}-k)
\int_{0}^{\frac{1}{2} \vert k-k_{F}\vert} dq\, 8q^2 \nonumber \\
&&+ \frac{1}{k} \int_{\frac{1}{2} \vert k-k_{F}\vert}^{\frac{1}{2}
( k+k_{F} )} dq \, q\left[ \left( k_{F}^2-q^2 \right) -4q\left( q-k_{F}\right)
\right] \biggr\rbrace \times \nonumber \\
&& \times \biggl[ \sum_{L'}\int_{0}^\infty dq'\, {q'}^2 \frac{ Q(
K_{av}, q' ) {\cal V}_{LL'}^{JST} (K_{av};q,q')^2}{\omega
-E_{2p1h}(K_{av},q, q') +i \eta} \biggr]\, . \label{eq:self2p1h}
\end{eqnarray}
The single-particle energies $\epsilon_{q}$ correspond to the
Hartree-Fock approximation for the single-particle energy
\begin{equation}
\epsilon_{q} = \frac{q^2}{2m} + \Sigma^{(HF)} (q) \, ,
\label{eq:hfsing}
\end{equation}
with $m$ for the mass of the nucleon.
In the second part of eq.(\ref{eq:self2p1h}) we have used the so-called
angle-averaged approximation for the Pauli operator, which is defined
by\cite{haftel}
\begin{equation}
Q(K,q) = \cases{0 ,& if $q \leq \sqrt{k_{F}^2 - K^2}$ \cr
{K^2 + q^2 - k_{F}^2 \over 2Kq} , & if $ \sqrt{k_{F}^2 - K^2} \leq q
\leq k_{F} + K$ \cr 1 , & if $q \geq k_{F} + K$ \cr} \label{eq:pauli}
\end{equation}
The other contribution to the second order self-energy with
intermediate 2-hole 1-particle states (2h1p), displayed in
Fig.\ref{fig:diag}c, can be calculated in a way very similar to
eq.(\ref{eq:self2p1h})
\begin{eqnarray}
\Sigma^{(2h1p)}(k,\omega ) & = & \sum\!\!\!\!\!\!\int_{p>F}
\sum\!\!\!\!\!\!\int_{h_{1},h_{2}<F} \frac {< kp \vert {\cal V}\vert
h_{1},h_{2}>^2}{\omega - (\epsilon_{h1}+\epsilon_{h2}-\epsilon_{p}) -
i\eta }\nonumber \\
& = &\sum_{LL'SJT}  \int_{0}^\infty dq\, W(q)
\int_{0}^\infty dq'\, {q'}^2 \frac{ P(
\tilde{K_{av}}, q' ) {\cal V}_{LL'}^{JST} (\tilde{K_{av}};q,q')^2}{\omega
-E_{2h1p}(\tilde{K_{av}},q, q') -i \eta}\, . \label{eq:self2h1p}
\end{eqnarray}
The Pauli operator $Q$, which ensured in eq.(\ref{eq:self2p1h}) that the
sum over intermediate 2-particle states is restricted to states above
the Fermi surface, is replace by a corresponding operator $P$ to ensure
that the intermediate 2-hole state are below the Fermi level
\begin{equation}
P(K,q) = \cases{0 ,& if $q \geq \sqrt{k_{F}^2 - K^2}$ \cr
{k_{F}^2- K^2 - q^2 \over 2Kq} , & if $ \sqrt{k_{F}^2 - K^2} \geq q
\geq k_{F} - K$ \cr 1 , & if $q \leq k_{F} - K$ \cr}
\label{eq:hpauli}
\end{equation}
The definitions of the mean value for the center of mass momentum
$\tilde{K_{av}}$ and the weight function $W(q)$ in the integral of
eq.(\ref{eq:self2h1p}) are a bit more involved than in the case of the
2p1h term and are given in the appendix\cite{knehr}.

After the definition of the self-energy we could now proceed and
calculate the corresponding single-particle Green's function by solving a
Dyson equation of the form\cite{review}
\begin{equation}
g(k,\omega ) = g^{(HF)} (k,\omega ) + g^{(HF)} (k,\omega ) \left[
\Sigma^{(2p1h)}(k,\omega ) + \Sigma^{(2h1p)}(k,\omega ) \right] g(k,
\omega)\, ,\label{eq:dyson}
\end{equation}
with the single-particle Green's function in the Hartree-Fock
approximation
\begin{equation}
g^{(HF)} (k,\omega ) = \frac{\Theta (k_{F}-k)}{\omega -\epsilon_{k}
-i\eta } + \frac{\Theta (k -k_{F})}{\omega -\epsilon_{k} + i\eta }\, .
\label{eq:greenhf}
\end{equation}
The momentum distribution $n(k)$ can then be calculated from the
imaginary part of the single-particle Green's function by
\begin{equation}
n(k) = \frac{1}{\pi} \int_{-\infty}^{\epsilon_{F}} d\omega\ \mbox{Imag}\
g(k, \omega ) \, . \label{eq:momen1}
\end{equation}

\subsection{Numerical Approach}
Instead of proceeding along the lines indicated in eqs.(\ref{eq:dyson})
- (\ref{eq:momen1}), we use the fact that in all numerical calculations
the integrals of eqs.(\ref{eq:self2p1h}) and (\ref{eq:self2h1p}) will
be discretized. This means that eq.(\ref{eq:self2p1h}) takes the form
\begin{equation}
\Sigma^{(2p1h)}(k,\omega ) = \sum_{i=1}^N \frac{F_{i}^2(k)}{\omega -
E_{i}^{(2p1h)}+i\eta}\, , \label{eq:simself1}
\end{equation}
while eq.(\ref{eq:self2p1h}) can be rewritten as
\begin{equation}
\Sigma^{(2h1p)}(k,\omega ) = \sum_{j=1}^M \frac{G_{j}^2(k)}{\omega -
E_{j}^{(2h1p)}-i\eta}\, . \label{eq:simself2}
\end{equation}
This discretization implies in particular that we represent the
singularities of the self-energy in terms of discrete poles slightly
above (2h1p) and below the real axes. This analytic structure of the
self-energy is identical to the one obtained for a finite system within
a model-space defined in terms of discrete single-particle states. This
means that we may use the same techniques to determine the features of
the single-particle Green's function as employed e.g. in
\cite{sko1,skouras}. Translating this technique into the present
example, this means that the single-particle Green's
function will be defined in the Lehmann representation by
\begin{equation}
g(k,\omega ) = \sum_{\alpha =1}^{N+M+1} \frac{X_{\alpha}^2}{\omega -
\omega_{\alpha} \pm i\eta }\, ,
\label{eq:lehman}
\end{equation}
with the sign in front of the infinitesimal imaginary part $\eta$
being positive for poles $\omega_{\alpha}$ above the Fermi energy
$E_{F}$ and negative else. For each $k$ the positions of these poles,
$\omega_{\alpha}$, and the residua, $X_{\alpha}^2$, can be determined
from the solution of the following eigenvalue problem
\begin{equation}
\pmatrix{\epsilon_{k} & F_1 & \ldots & F_N & G_1 & \ldots &
G_M \cr
F_1 & E_1^{(2p1h)} & & 0 & & & \cr \vdots & & \ddots & & &  & \cr
F_{N} & 0 & & E_N^{(2p1h)} & & & 0 \cr G_1 &  & & &  E_1^{(2h1p)} & & \cr
\vdots & & & &
& \ddots & \cr G_{M} & 0 & \ldots & 0 & & \ldots & E_M^{(2h1p)} \cr }
\pmatrix{ X_{\alpha} \cr Y_{\alpha,1} \cr \vdots
\cr Y_{\alpha,N} \cr Z_{\alpha,1}\cr \vdots \cr Z_{\alpha,M}
\cr }
= \omega_\alpha
\pmatrix{ X_{\alpha} \cr Y_{\alpha,1} \cr \vdots
\cr Y_{\alpha,N} \cr Z_{\alpha,1}\cr \vdots \cr Z_{\alpha,M}
\cr } \; .
\label{eq:matr1}
\end{equation}
Note that the dimension of this matrix ($N+M+1$) as well as the matrix
elements $F_{i}$, $E_{i}^{(2p1h)}$ and $G_{j}$, $E_{j}^{(2h1p)}$ refer
to the nomenclature employed in eqs.(\ref{eq:simself1}) and
(\ref{eq:simself2}), respectively. Using the representation of the
Green's function in eq.(\ref{eq:lehman}) the occupation probabilities
are calculated easily as
\begin{equation}
\tilde n(k) = \sum_{\alpha} \Theta(E_{F} - \omega_{\alpha}) X_{\alpha}^2
\, , \label{eq:momen2}
\end{equation}
which leads to the momentum distribution if we divide by the
single-particle density
\begin{equation}
n(k) = \frac{\tilde n(k)}{\rho} = \frac{3 \tilde n(k)}{4 \pi k_{F}^3}
\, ,\label{eq:momen3}
\end{equation}
with $k_{F}$ the Fermi momentum of the nuclear matter system.

\subsection{BAGEL Approximation}
Using the Lehmann representation of the single-particle Green's
function of eq.(\ref{eq:lehman}) the continua of states of
nuclear matter with one additional nucleon and one hole are represented
in terms of some discrete energies $\omega_{\alpha}$. Depending on the
accuracy of the discretization on eqs.(\ref{eq:simself1}) and
(\ref{eq:simself2}) the number of eigenvalues typically considered in
numerical calculations ranges from a few hundred up to a few thousand.
This may be compared to the Hartree-Fock approximation of
eq.(\ref{eq:greenhf}), in which the Green's function for a nucleon with
momentum $k$ is represented by just one pole.
This number of pole terms is not a problem as long as one is just
interested in the evaluation of the Green's function or simple
observables as the momentum distribution. The structure of the Green's
function in eq.(\ref{eq:lehman}), however, may be too complicate
to be used in the evaluation of quantities which are defined in terms
of products of these Green's functions. Examples for such quantities
are e.g.~the various response functions of nuclear matter or a
self-consistent evaluation of the self-energy, which implies that the
self-energies to be used in the Dyson equation (\ref{eq:dyson}) are
calculated in terms of the resulting Green's functions. For such
calculations it may be preferable to ``optimize'' the number of pole
terms in eq.(\ref{eq:lehman}), which means: Try to find a minimum
number of poles, which yields the same observables than the complete
Green's function.

In order to develop such an efficient representation of the Green's
function we try to apply the so-called ``BAsis GEnerated by Lanczos"
(BAGEL) scheme, which has successfully been used for the description of
finite nuclei in finite model spaces\cite{kuo,sko1,skouras}.
For that purpose we consider the operator $\hat a$ which corresponds to
a part of the matrix in eq.~(\ref{eq:matr1})
\begin{equation}
\hat a =
\pmatrix{\epsilon_k & F_1  &\ldots & F_N \cr
F_1 & E_{1}^{(2p1h)} & &  \cr \vdots &  & \ddots &  \cr F_{N} &  &
&E_{N}^{(2p1h)} \cr} \; ,
\label{eq:bag1}
\end{equation}
and apply this operator on the single-particle state $\vert \alpha >$, which
in terms of the matrix representation of eq.~(\ref{eq:bag1}) is described
by the column vector $ (1, 0 \dots 0)^T$
\begin{equation}
\hat a \vert \alpha > = \epsilon_{k} \vert \alpha > + \tilde
a_1 \vert \alpha_1 > \, ,
\end{equation}
where $\vert \alpha_1 >$ is orthogonal to $\vert \alpha >$ and the coefficient
$\tilde a_1$ is chosen so that $\vert \alpha_1 >$ is  normalized.
Following the Lanczos algorithm \cite{lanczos}, one can subsequently
construct additional states $\vert \alpha_{i} >$, which are all
orthogonal to each other. Applying the Lanczos procedure $n$ times one
obtains $n$ basis states of the 2p1h configuration space.

In a similar way we can furthermore construct $m$ basis states of the
2h1p configuration space by
considering the corresponding sub matrix of eq.(\ref{eq:matr1})
\begin{equation}
\hat A =
\pmatrix{\epsilon_k & G_1  &\ldots & G_N \cr
G_1 & E_{1}^{(2h1p)} & &  \cr \vdots &  & \ddots &  \cr G_{M} &  &
&E_{M}^{(2h1p)} \cr} \;
,\label{eq:bag3}
\end{equation}
and reduce the eigenvalue problem of eq.(\ref{eq:matr1}) to the
corresponding one in the subspace defined by the basis of the single-particle
state plus the $(n+m)$ basis states generated by the Lanczos scheme
just outlined. The Green's function of this BAGEL(n,m) approximation is then
defined according to eq.(\ref{eq:lehman}) using the $(n+m+1)$
eigenvalues and vectors obtained from the diagonalization of the matrix
truncated to the subspace. It is obvious
that the BAGEL(0,0) corresponds to the HF approximation, while for $n$
approaching $N$ and $m$ close to $M$ the BAGEL(n,m) approximation for
the Green's function becomes identical to
the exact solution of eqs.~(\ref{eq:matr1}) and (\ref{eq:lehman}).

\subsection{Effective Interaction}
At the end of this section we want to define the effective NN interaction
${\cal V}$, used in the definition of the self-energy above. One
possible choice would be of course to replace ${\cal V}$ by the bare NN
interaction. As it has been discussed already in the introduction, the
HF approximation in terms of a realistic NN interaction is not a very
useful approach and it is not clear whether a perturbation expansion in
terms of the bare NN interaction up to second order, as just outlined, will
be sufficient. Therefore we employ the $G$ matrix, an appropriate
solution of the Bethe-Goldstone equation for ${\cal V}$. The starting
energy $Z$ in the Bethe-Goldstone equation is chosen according to the
Brueckner-Hartree-Fock (BHF) choice for the self-energy of a nucleon with
momentum $k$ below the Fermi momentum and put to be the average
of two single-particle states below the Fermi energy if $k$ is above
the Fermi momentum. This ensures
that $G$ remains real. With this choice  we employ an approach
which is very similar to the one used for finite nuclei in \cite{pap2},
where the self-energy is also calculated including terms up to second
order in a nuclear matter $G$-matrix.

Using the $G$ matrix for the effective interaction also implies,
however, that we have to face a double-counting problem. The diagram of
second order, displayed in fig.\ref{fig:diag}b is to some extent
already taken into account in the Brueckner-Hartree-Fock (BHF) approach
for the self-energy displayed in fig.\ref{fig:diag}a. This
double-counting does not directly effect the calculation of the
momentum distribution. The choice for the starting energy just
presented leads to a real self-energy contribution of
eq.(\ref{eq:selfhf}) without any poles and therefore the momentum
distribution calculated for a self-energy, which only accounts for this
term, remains identical to the HF one. There is a self-consistency
problem, however, with respect to the energy spectrum reflected in the
poles of the Green's function (see eq.(\ref{eq:lehman})). If for the
moment we ignore the 2h1p contribution to the self-energy and evaluate
the Green's function according to scheme outlined in
eqs.(\ref{eq:lehman}) - (\ref{eq:matr1}) for a nucleon with momentum
below $k_{F}$, we will find one eigenstate of eq.(\ref{eq:matr1}) with
negative energy and a large coefficient $X_{\alpha}$, the quasihole
state, and $N$ eigenvalues at positive energies. Due to the
diagonalization, however, the energy of the quasihole state,
$\epsilon_{k}^{(qh,2p1h)}$, will be substantially below the corresponding HF
energy $\epsilon_{k}$. Therefore we replace the first element of the
matrix in eq.(\ref{eq:matr1}) by
\begin{equation}
\epsilon_{k}\, \Longrightarrow\, \tilde\epsilon_{k} =
\epsilon_{k}-(\epsilon_{k}^{(qh,2p1h)} - \epsilon_{k})\, .
\label{eq:shift}
\end{equation}
This shift in energy ensures that the quasihole state for a self-energy
with inclusion of only the 2p1h term will essentially be identical to
the BHF energy. Therefore the double-counting is removed. Note again that
this double-counting problem does not affect the calculation of the
momentum distribution. The energy-shift is useful, however, to obtain a
realistic energy spectrum for the poles of the Green's functions.

\section{Momentum Distribution in Finite Nuclei}

\subsection{Direct Approach}
\label{sec3.1}
Also the calculation of the momentum distribution presented in
\cite{pap1,pap2} directly for the nucleus $^{16}O$ is based on a self-energy
of the nucleon calculated up to second  order in a nuclear matter
$G$-matrix as described by the diagrams of fig.\ref{fig:diag}.
As a first step one considers the HF contribution to the self-energy
\begin{equation}
\Sigma^{HF}_{l_1j_1} (k_1,k'_1) =
\frac{1}{2(2j_1+1)} \sum_{n_2 l_2 j_2 J T} (2J+1) (2T+1)
\left\langle k_1 l_1 j_1 n_2 l_2 j_2 J T \right| G \left|
k'_1 l_1 j_1 n_2 l_2 j_2 J T\right\rangle .
\label{eq:selhf}
\end{equation}
The matrix elements of $G$ used in this expression are antisymmetrized NN
matrix elements calculated in the laboratory system
The quantum numbers $l_{i}$ and
$j_{i}$ refer to the orbital and total angular momentum of the single
nucleons in this frame and $J$ and $T$ denote the angular momentum and
isospin of the 2-particle states. The matrix elements are calculated in
a mixed representation with $n_{i}$ referring to the radial quantum
numbers of oscillator bound (hole) states, whereas the $k_{i}$
denote the absolute value of the momentum for a free particle state.
The summation over the oscillator quantum numbers is restricted to the
states occupied in the independent particle model  of $^{16}$O. This
Hartree-Fock part of the self-energy is real and does not depend on the
energy. The HF single-particle wave functions can be obtained by
expanding them
\begin{equation}
\vert \alpha^{HF} ljm > = \sum_{i} \vert K_{i} ljm >
< K_{i} \vert \alpha^{HF}  >_{lj}
\label{eq:hfexpan}
\end{equation}
in a complete and orthonormal set of regular basis
functions within a spherical box of radius $R_{\rm box}$ which is large as
compared to the radius of the nucleus
\begin{equation}
\Phi_{iljm} ({\bf r}) = \left\langle {\bf r} \vert K_i l j m
\right\rangle = N_{il} j_l(K_ir)
{\cal Y}_{ljm} (\vartheta\varphi ) \label{eq:boxbas}
\end{equation}
In this equation ${\cal Y}_{ljm}$ represent the spherical harmonics
including the spin degrees of freedom and $j_l$ denote the spherical
Bessel functions for the discrete momenta $K_i$ which fulfill
\begin{equation}
j_l (K_i R_{\rm box}) = 0 .
\label{eq:bound}
\end{equation}
Using the normalization constants
\begin{equation}
N_{il} =\cases{\frac{\sqrt{2}}{\sqrt{R_{\rm box}^3} j_{l-1}(K_i
R_{\rm box})}, & for $l > 0$ \cr
\frac{i \pi\sqrt{2}}{\sqrt{R_{\rm box}^3}}, & for $l=0$,\cr}
\label{eq:nbox}
\end{equation}
the basis functions defined in Eq. (\ref{eq:boxbas}) are orthogonal and
normalized within the box. The expansion coefficients of
eq.(\ref{eq:hfexpan}) are obtained by diagonalizing the HF Hamiltonian
\begin{equation}
\sum_{n=1}^{N_{\rm max}} \left\langle K_i \right| \frac{K_i^2}{2m}\delta_{in} +
\Sigma^{HF}_{lj} \left| K_n \right\rangle \left\langle K_n \vert
\alpha^{HF} \right\rangle_{lj} = \epsilon^{HF}_{\alpha
lj} \left\langle K_i \vert \alpha^{HF}\right\rangle_{lj}. \label{eq:hfequ}
\end{equation}
Here and in the following the set of basis states in the box has been
truncated by assuming an appropriate $N_{\rm max}$. From the HF wave
functions and energies one can construct the HF approximation to the
single-particle Green's function in the box (compare eq.(\ref{eq:greenhf})),
which for the finite nucleus has the form
\begin{equation}
g_{\alpha lj}^{(HF)} (k_{i}, k_{j}; \omega ) = \frac{<k_{i}\vert
\alpha^{HF} >_{lj}<\alpha^{HF}\vert
k_{j}>_{lj}}{\omega -\epsilon^{HF}_{\alpha lj} \pm
i\eta} , \label{eq:green0}
\end{equation}
As an example for the contributions to the self-energy of second order
in $G$ we recall the calculation of the 2p1h term. In the approach of
\cite{pap2} one first calculates the imaginary part of this self-energy
contribution, depending on the energy $\omega$
\begin{eqnarray}
{W}^{2p1h}_{l_1j_1} (k_1,k'_1; \omega )
=& \frac{-1}{2(2j_1+1)}  \sum_{n_2 l_2 j_2} \sum_{l_3 l_4 j_3 j_4}
\sum_{J T} \int k_3^2 dk_3 \int k_4^2 dk_4  (2J+1) (2T+1) \nonumber \\
& \times \left\langle k_1 l_1 j_1 n_2 l_2 j_2 J T \right| G \left|
k_3 l_3 j_3 k_4 l_4 j_4 J T \right\rangle \nonumber \\
& \times  \left\langle k_3 l_3 j_3 k_4 l_4 j_4 J T \right| G \left|
  k'_1 l_1 j_1 n_2 l_2 j_2 J T
  \right\rangle \nonumber \\
& \times \pi \delta\left(\omega +\epsilon_{n_2 l_2 j_2}
-\frac{k_3^2}{2m}-\frac{k_4^2}{2m}\right), \label{eq:w2p1h}
\end{eqnarray}
where the ``experimental'' single-particle energies $\epsilon_{n_2 l_2 j_2}$
are used for the hole states (-47 MeV, -21.8 MeV, -15.7 MeV for $s_{1/2}$,
$p_{3/2}$ and $p_{1/2}$ states, respectively), while the energies of the
particle states are given in terms of the kinetic energy only.
The expression in Eq. (\ref{eq:w2p1h}) still ignores the requirement that the
intermediate particle states must  be orthogonal to the hole states, which
are occupied for the nucleus under consideration. The techniques to
incorporate the orthogonalization of the intermediate plane wave states
to the occupied hole states as discussed in detail by Borromeo et
al.\cite{boro} have also been used here.
The 2h1p contribution to the imaginary part ${W}^{2h1p}_{l_1j_1}(k_1,k'_1;
\omega )$ can be calculated in a similar way (see also \cite{boro}).

The choice to assume pure kinetic energies for the particle states in
calculating the imaginary parts of $W^{2p1h}$ (Eq. (\ref{eq:w2p1h})) and
$W^{2h1p}$ may not be very realistic for the excitation modes at low energy.
Indeed a seizable imaginary part in $W^{2h1p}$ is obtained only for
energies $\omega$ below -40 MeV. As we are mainly interested, however, in the
effects of short-range correlations, which lead to excitations of particle
states with high momentum, the choice seems to be appropriate.
A different approach would be required to treat the coupling
to the very low-lying two-particle-one-hole and two-hole-one-particle
states in an adequate way. Attempts at such a treatment can be found in
Refs.\ \cite{brand,rijsd,skou1}.

The real parts of the 2p1h and 2h1p terms in the self-energy can be
calculated from the corresponding imaginary parts by using dispersion
relations\cite{mahau}. As an example we present the dispersion relation
for the  2p1h part, which is given by
\begin{equation}
V^{2p1h}_{l_1j_1}(k_1,k_1';\omega )=\frac{P}{\pi} \int_{-\infty}^{\infty}
\frac {W^{2p1h}_{l_1j_1}(k_1,k_1';\omega ')}{\omega '-\omega} d\omega ',
\label{eq:disper1}
\end{equation}
where $P$ means a principal value integral. Putting the various
contributions together the correction to the HF self-energy due to the
second order terms can be written
\begin{equation}
\Delta\Sigma_{lj}(k_{1},k_{2};\omega ) = \bigl( V^{2p1h} - V_{c} +
V^{2h1p}\bigr) + i\bigl( W^{2p1h} + W^{2h1p} \bigr)
\end{equation}
where $V_{c}$ denotes a correction term to account for double-counting
between the $V^{2p1h}$ and ladder contributions already contained in
the HF part of the self-energy\cite{pap2} (see also discussion at the
end of section 2). With this correction to the self-energy one can
solve a Dyson equation for the complete Green's function (see also
eq.(\ref{eq:dyson})), which corresponds to an integral equation for finite
systems
\begin{equation}
  g_{lj}(k_1,k_2;\omega ) = g^{(HF)}_{lj}(k_1,k_2;\omega )
+ \int dk_3\int dk_4 g^{(HF)}_{lj}(k_1,k_3;\omega ) \Delta\Sigma_{lj}
(k_3,k_4;\omega )  g_{lj}(k_4,k_2;\omega ) \,.
\label{eq:dyson2}
\end{equation}
{}From the imaginary part of this Green's function one can finally
evaluate the momentum distribution according to
\begin{equation}
n(k) = \sum_{lj} 2(2j+1) \int_{-\infty}^{\epsilon_{F}} d\omega \,
\frac{1}{\pi} {\rm Imag} \bigl[g_{lj}(k,k;\omega )\bigr] \, .
\label{eq:momfin}
\end{equation}

\subsection{Local Density Approximation}
\label{sec3.2}
Instead of evaluating the momentum distribution directly for the finite
nucleus one can try to deduce the effects of correlations on the
momentum distribution from the investigation of nuclear matter. As a
first step towards such an approach we consider the local density
$\rho^{\rm HF} (r)$ and the momentum distribution $n^{\rm HF} (k)$
derived from the solution of the Hartree-Fock eq.(\ref{eq:hfequ}).
Using this density distribution we can define a local Fermi momentum
\begin{equation}
k_{F}^{\rm local} (r) = \left[ \frac{3 \pi^2 \rho^{\rm HF}
(r)}{2}\right]^{1/3}
\label{eq:kfloc}
\end{equation}
and evaluate an average occupation number for the states occupied in
the mean field approach by
\begin{equation}
N_{\rm aver}= \frac{4 \pi}{A}  \int r^2 \, dr \, \rho^{\rm HF} (r)
\bar{n} (k_{F}^{\rm local}(r)) \, ,
\label{eq:naver}
\end{equation}
with $A$ the number of nucleons ($A$=16 in our example of $^{16}$O) and
$\bar n$ the occupation number (see eq.(\ref{eq:momen2})) calculated
for nuclear matter with the local Fermi momentum and averaged over all
momenta below this Fermi momentum. With this average occupation number
one can account for the depletion of the occupation of states occupied
in HF approximation. The high momentum components originating from the
partial occupation of states above the Fermi momentum are then
evaluated as
\begin{equation}
\Delta n (k) = 4 \pi k^2 \, \int 4\pi r^2 \, dr \, \Theta\left( k-
k_{F}^{\rm local}(r)\right) n \left(k; k_{F}^{\rm local}(r)\right)\,
\label{eq:deltan}
\end{equation}
where $n(k;k_{F})$ is the momentum distribution of nuclear matter
according to eq.(\ref{eq:momen3}) calculated for the local Fermi
momentum. The total momentum distribution is then given as
\begin{equation}
n^{LDA}(k) =  N_{\rm aver}n^{\rm HF} (k) + \Delta n (k)\, .
\label{eq:momen4}
\end{equation}
Within this local density approximation we also would like to estimate
the spectral strength which is missing in the calculation according to
eq.(\ref{eq:momfin}) due to a restriction in the sum of that equation
to partial waves with orbital angular momentum up to $l=L_{max}$. For
our LDA approximation that restriction would mean to consider
contributions to $\Delta n(k)$ in eq.(\ref{eq:deltan}) with
\begin{equation}
\vert \vec l \vert = rk\sin{\varphi} \leq \hbar
\sqrt{L_{max}(L_{max}+1)} \, ,
\end{equation}
$\varphi$ denoting the angle between $r$ and the momentum $k$. This
means that the integrand in eq.(\ref{eq:deltan}) should be reduced by a
factor
\begin{equation}
\frac{2}{\pi} \arcsin{\frac{\hbar\sqrt{L_{max}(L_{max}+1)}}{rk}}\, .
\label{eq:lmax}
\end{equation}

\section{Results and Discussion}

\subsection{Nuclear Matter}
In order to evaluate the momentum distribution for nuclear matter at a
given density $\rho$, which may as well be characterized by the
corresponding Fermi momentum $k_{F}$, using
the method outlined in the previous section, we have to determine as a
first step the spectrum of HF single-particle energies $\epsilon_{k}$.
In our approach, defining the HF contribution to the self-energy (see
\ref{fig:diag}b) in terms of the nuclear matter G-matrix, these
single-particle energies correspond to the single-particle energies
obtained in the BHF approximation. For the OBE potential $B$ of
\cite{rupr} the BHF single-particle energies have been
parameterized\cite{rolf} in terms of an effective mass $m^*$ and a
constant shift $C$ by
\begin{equation}
\epsilon_{k} = \sqrt{k^2 + {m^*}^2} - m^* + m + C\, .
\label{eq:paraeps}
\end{equation}
The parameters $m^*$ and $C$ as a function of the Fermi momentum $k_{F}$
are listed in table 2 of \cite{rolf}. It should be mentioned that in
our study we have used the so-called non-relativistic parameterization
since in our present study we ignore all effects of the Dirac BHF
approach due to a change of the Dirac spinors of the nucleons in the
nuclear medium. With this definition of the single-particle spectrum
one can evaluate the matrix in eq.(\ref{eq:matr1}) with the
renormalization of eq.(\ref{eq:shift}), solve the eigenvalue problem of
eq.(\ref{eq:matr1}) and determine the momentum distribution with
eq.(\ref{eq:momen2}).

Results for this momentum distribution in nuclear
matter at the empirical saturation density ($k_{F}$=1.36 fm$^{-1}$) are
displayed in Fig.\ref{fig:nm1}. The momentum distribution derived from
the OBE potential (dashed-dotted line) is compared to the one obtained
using the Reid soft-core potential employing the same technique (dashed
line). The Reid soft-core potential predicts stronger effects of
correlations in this momentum distribution. This is characterized by a
stronger depletion of the states with momenta $k$ below the Fermi
momentum (the Reid soft-core yields an average occupation of these
states of 0.83  while the OBE predicts 0.86 ) as well as larger probability
at higher momenta (for $k \approx$ 4 fm$^{-1}$ the density which is obtained
for the Reid potential is by more than a factor 2 larger as the one
deduced from the OBE). This is in agreement with the observation that
the modern OBE potentials are ``softer'' and contain a weaker tensor
force as it is also reflected in the D-state probability calculated for
the deuteron\cite{rupr}.

Figure \ref{fig:nm1} also shows the prediction for the momentum
distribution of nuclear matter obtained for the Reid soft-core
potential using the much more sophisticated techniques of \cite{vonde}.
The good agreement of our approach, in which the self-energy is
calculated in a perturbative scheme, including terms up to second order
in $G$, with the one of \cite{vonde} where the particle-particle
hole-hole ladders are taken into account to all order using a
self-consistent single-particle Greens function, gives us some
confidence that the present approach provides reliable information for
systematic studies in nuclear matter as well as finite nuclei.

The sensitivity of the calculated momentum distribution on the nuclear
density is demonstrated in fig.\ref{fig:nm2}. In order to allow a
direct comparison this figure does not show the momentum distribution
but the occupation of the single-particle states as a function of the
momentum $k$ in units of the Fermi momentum ($k/k_{F}$). The results
for the occupation of states below the Fermi momentum ($k/k_{F}\leq 1$)
are shown with respect to the linear scale on the left axis, while the
occupation of states with momenta larger than $k_{F}$ are shown with
respect to the logarithmic scale on the axis at the right hand side.
One observes that the calculated occupations are rather insensitive to
density of the nuclear system. Only for very small densities
($k_{F}$=0.8 fm$^{-1}$, which corresponds to roughly 20 percent of the
empirical saturation density) one finds occupation probabilities which
are considerably smaller. This might be an indication of the
instability of homogeneous nuclear matter at such small
densities\cite{neck,benh}.

In order to explore the density dependence a bit more in detail, we
have separated the contributions to the single-particle
density into a quasihole contribution and a continuum contribution. For
momenta $k$ below the Fermi momentum a large contribution to the
momentum distribution of eq.(\ref{eq:momen2}) originates from one
eigenstate $\alpha$ with a maximal coefficient $X_{\alpha}$ and
an eigenvalue $\omega_{\alpha}$ which we
identify as the quasihole energy $E_{qh}(k)$. In particular at small
momenta ($k \approx 0.2 k_{F}$) one also finds that a few states around
the quasihole energy exhibit large coefficients $X_{\alpha}$ and
therefore contribute significantly to the sum in eq.(\ref{eq:momen2}).
We define the quasihole strength to be the contribution of all terms in
eq.(\ref{eq:momen2}) which originate from an eigenstate of
eq.(\ref{eq:matr1}) with an eigenvalue within an interval of length 3
MeV around the quasihole energy $E_{qh}$
\begin{equation}
 N_{qh}(k) = \sum_{\alpha} \Theta\bigl(E_{F} - \omega_{\alpha}\bigr)\
\Theta\bigl(E_{qh}(k) +1.5 - \omega_{\alpha}\bigr)\ \Theta\bigl(
\omega_{\alpha}-E_{qh}(k)+1.5\bigr)\  X_{\alpha}^2
\, . \label{eq:nqh}
\end{equation}
For momenta $k$ larger than the Fermi momentum the eigenvalue of the
state with maximal expansion coefficient $X_{\alpha}$ occurs at
energies above the Fermi energy $E_{F}$ and therefore we don not obtain
any quasihole strength for those momenta. The quasihole strength
$N_{qh}$ is shown in fig.\ref{fig:nm3} as a function of the ratio
$k/k_{F}$ for two densities (lines labeled with triangles). The
remaining contributions to the sum in  eq.(\ref{eq:momen2}) will be
called the continuum contribution to the occupation probability or
momentum distribution.

{}From the inspection of the results displayed in fig.\ref{fig:nm3} one
finds that the quasihole contribution to the occupation probability
increases drastically with the momentum while the continuum
contribution decreases in a corresponding way. A typical ratio of
quasihole versus continuum contribution is 0.6 for small momenta but as
large as 10 for momenta close to the Fermi momentum. This means that
the energy distribution of the single-particle strength is highly localized
at the quasihole energy for states with momenta close to $k_{F}$
whereas one observes a brought distribution and fragmentation of the
strength for small momenta, i.e.~deeply bound hole states.

The continuum contribution to the occupation probability decreases
monotonically with increasing momentum and is a rather smooth
function even at
the Fermi momentum. This implies that the gap in the momentum
distribution exhibited e.g. in fig.\ref{fig:nm1} at $k=k_{F}$
originates simply from the fact that the quasihole contribution
vanishes since the energy of the corresponding state gets larger than
the Fermi energy. It is worth noting that about 70 percent of the
single-particle strength is located in the quasihole contribution. Two
thirds of the remaining continuum contribution occurs at momenta below
the Fermi momentum and only one third of the continuum contribution,
which means slightly more than 10 percent of the total strength occurs
at momenta above $k_{F}$.

{}From the single-particle Green's function we also determine
the mean value of the
energy for the spectral distribution at a given momentum
\begin{equation}
\bar \omega (k) = \frac{\frac{1}{\pi} \int_{-\infty}^{\epsilon_{F}} d\omega\
\omega \mbox{Imag}\ g(k, \omega )}{n(k)} \, , \label{eq:wbar1}
\end{equation}
or translated into the tools we are using in our numerical treatment
\begin{equation}
\bar \omega (k) = \frac{1}{\tilde n(k)} \sum_{\alpha} \omega_{\alpha}
\Theta(E_{F} - \omega_{\alpha}) X_{\alpha}^2
\, , \label{eq:barom}
\end{equation}
with the occupation number $\tilde n$ calculated following
eq(\ref{eq:momen2}). The summation in this equation can be truncated
as discussed above to determine the mean value for the energy resulting
from the continuum part of the momentum distribution.

Such mean values are presented in the right part of fig.\ref{fig:nm4}.
One finds that these mean values for the continuum are more negative
than the corresponding HF single-particle energies or the energies of
the quasihole states, which are shown in the left part of fig.\ref{fig:nm4}
(note the different scales on the axes). Particularly at large momenta,
above $k_{F}$ where the continuum part represents the total momentum
distribution, these mean values $\bar \omega (k)$ are very attractive.
This implies that these high momentum components of the momentum
distribution occur predominantly at large excitation energies of the
residual nuclear system, which corresponds to large missing energies in
knock out experiments. This result for nuclear matter confirms the
observations made for finite systems in \cite{pap1,pap2}.

In the left part of fig.\ref{fig:nm4} the BHF single-particle energies
(lines labeled with triangles) are compared to the energies $E_{qh}$ of the
quasihole energies. One can see that the inclusion of the 2h1p terms of
fig.\ref{fig:diag}c yields a repulsive contribution to the quasihole
energy. This is specially true for states with momenta well below the
Fermi momentum. This means that the removal energy for nucleon knock
out experiments exciting states with large spectroscopic factor should
be much smaller for these deeply bound states than predicted in BHF
calculations.

As a last point in this subsection we would like to explore the
sensitivity of the calculated momentum distribution on the HF
single-particle spectrum. For that purpose we have modified the
effective mass parameter $m^*$ in the parameterization of
eq.(\ref{eq:paraeps}) from the BHF value of $m^*$=623 MeV \cite{rolf}
at $k_{F}$ = 1.35 fm$^{-1}$ by $\pm$ 100 MeV. From the upper part of
fig.\ref{fig:nm5} one can see that a reduction of the effective mass,
i.e.~the single-particle energy shows a stronger momentum dependence,
yields a reduction of the correlation effect. This reduction of the
correlations is indicated by an enhancement of the occupation of states
with $k\leq k_{F}$ and a reduction of the high momentum components.
Another modification of the HF single-particle spectrum can be obtained
by introducing a gap between the energies of particle and hole states.
As on can observe from the lower part of fig.\ref{fig:nm5} such a gap
does not affect the high momentum components very much but only the
occupation probability around $k_{F}$.

\subsection{BAGEL Approximation}
In order to investigate the efficiency of the BAGEL approximation
introduced in the previous section for the representation of the
single-particle Green's function in terms of a few poles we have
evaluated the momentum distribution of nuclear matter considering
various combinations BAGEL(n,m) for the number of basis states $n$
in the 2p1h and $m$ in the 2h1p part of the eigenvalue problem
eq.(\ref{eq:matr1}). Results for a few examples are displayed in
fig.\ref{fig:nm6}.

One finds that a very good approximation for the high-momentum
components is obtained already with a very small number of basis
states. The occupation probabilities for $k > k_{F}$ are reproduced in
quite a satisfactory way in the simplest approximation BAGEL(1,1) and
the results become indistinguishable from the exact results if we use
any approximation with $n,m$ larger than 1. The convergence of the
BAGEL approximation towards the exact result with increasing $n,m$ is
not as good for the occupation of states below $k_{F}$. A larger number
of pole terms is required in particular to reproduce the decrease of
the occupation number with $k$ getting close to $k_{F}$.

Of course it is still a very efficient approximation to reduce the
number of poles in the Lehmann representation of the single-particle
Green's function of eq.(\ref{eq:lehman}) from a few hundred obtained by
an optimized discretization of the integrals in eqs.(\ref{eq:self2p1h})
and (\ref{eq:self2h1p}) to $n+m+1$ = 21 in the BAGEL(15,5)
approximation, but a closer inspection may help us to reduce the number
of terms even more.

Analyzing the basis states which are generated  by the BAGEL approach
in the 2p1h part
of the Hilbert space by applying $\hat a$ of eq.(\ref{eq:bag1})
according to the Lanczos scheme, one observes that states are generated
with very large eigenvalue $\omega_{\alpha}$ but negligible amplitude
$X_{\alpha}$. It requires some iteration steps to generate a few states
with lower energy and non-negligible coefficient. This generation of
basis states with extreme energies is of course a feature of the
Lanczos approach, which is not optimal for our present purpose. The
situation is better for the generation of basis states in the 2h1p sector, as
the eigenvalues of $\hat A$ (see eq.\ref{eq:bag1})) are more limited.

{}From this discussion we see that the BAGEL approximation for the
Green's function can be made more efficient by either ignoring the
contributions of those poles in the Lehmann representation of
eq.(\ref{eq:lehman}), which show very small coefficients $X_{\alpha}$
or to replace the Lanczos algorithm of generating the basis states by
one, which preferably generates eigenstates close to the Fermi energy.

\subsection{Finite Nuclei}
As a first example for the momentum distribution calculated directly
for the finite nucleus $^{16}O$ using the method described in
\cite{pap2} and briefly reviewed in section \ref{sec3.1}, we present in
fig.\ref{fig:o1} as a typical example the momentum distribution
obtained for $p_{1/2}$ partial wave. This refers to the corresponding
contribution to the sum in eq.(\ref{eq:momfin}) without the factor
$2(2j+1)$ for degeneracy of these states. As in \cite{pap2} we split
the momentum distribution into a quasihole contribution, which should
be observed if the $(A-1)$ nucleus remains in its groundstate, and a
continuum part reflecting the momentum distribution observed at larger
missing energies.

The results obtained for the Reid soft-core potential\cite{reid}
(dashed lines) are rather similar to those evaluated for the OBE
potential $B$ of \cite{rupr} (solid lines). However, there are some
characteristic differences which can also be observed in the other
partial waves: The quasihole contribution evaluated for the Reid
potential exhibits a maximum at smaller momenta and drops faster with
increasing momentum. This reflects the fact that nuclear structure
calculations like BHF yield less binding energy and a larger radius
using the Reid soft-core potential as compared to the OBE model for the
NN interaction. The continuum part, on the other side, exhibits larger
contributions at high momenta using the Reid potential. Following the
arguments present in the introduction of this paper, this would be an
indication that the Reid potential predicts "stronger" correlations.

Another indicator for the importance of NN correlations are the
occupation probabilities for the various partial waves as they are
listed in table \ref{tab:tab1}. These occupation probabilities are
obtained by a momentum integration of the various partial wave contributions
in eq.(\ref{eq:momfin}). Again we distinguish between quasihole
and continuum contribution for the partial waves with $l \leq 1$ and
compare the results for the two models of the NN interaction. The
results presented here for the OBE potential deviate slightly from
those presented in \cite{pap2} as we have increased the interval for
the energy integration in eq.(\ref{eq:momfin}) in our present study.
It is interesting to note that the occupation probabilities are quite
similar for both interactions for the orbits with angular momentum $l
\leq 2$ with slightly larger values for the OBE potential.
For the partial waves with larger $l$, however, the Reid
potential predicts occupation probabilities which are significantly
larger than those for the OBE potential. This difference seems to be
due to the stronger tensor component contained in the Reid potential.

Multiplying the occupation probabilities of table \ref{tab:tab1} with
the degeneracy factors $2(2j+1)$ one finds that 2.05 (2.09, using Reid)
``nucleons'' out of the 16 for $^{16}O$ are represented by the
continuum part of the momentum distribution. The total nucleon numbers,
including the quasihole part are 16.07 and 15.96 for the OBE and Reid
potential, respectively. This means that the particle-number violating
features of our present approach lacking a self-consistent treatment of
the single-particle Green's function are not very strong \cite{review}.

The continuum part of the total momentum distribution including
partial waves with $l \leq 4$ in the summation of eq.(\ref{eq:momfin})
is displayed in the left part of fig. \ref{fig:o2}. Again we can
observe the characteristic differences obtained for the two
interactions: While the OBE potential yields a momentum distribution
which is slightly larger at small momenta, the Reid potential predicts
contributions which are larger by almost a factor of 2 at large
momenta. This is in complete agreement with the observation made above
for nuclear matter (see fig.\ref{fig:nm1}).

Figure \ref{fig:o2} also exhibits the spectral distribution of the
continuum part. This spectral distribution is obtained by replacing the
energy integration in eq.(\ref{eq:momfin}) by a momentum integration.
One observes that the spectral distribution derived for the OBE
potential is slightly larger at energies with small absolute value,
which corresponds to small missing energies, while the Reid potential
yields a larger result at large missing energies. Concluding we may
characterize the differences between the two interactions by the
statement: The Reid potential predicts a larger component of the
single-particle density at large momenta and large missing energies as
compared to OBEP $B$.

Finally, we would like to discuss the validity of the local density
approximation (LDA) introduced in section \ref{sec3.2} for the description of
the momentum distribution at large momenta. Since this high-momentum
part of the momentum distribution is dominated by the continuum
contribution $\Delta n(k)$ in eq.(\ref{eq:momen4}) we will restrict our
discussion to this part and its comparison to the continuum part
of the momentum distribution evaluated directly for finite nuclei.

The result for $\Delta n(k)$ calculated according to
eq.(\ref{eq:deltan}) using the OBE interaction is represented by the
solid line in the left part of fig.\ref{fig:o3}. In the discussion of
the LDA above we already introduced a scheme to simulate the effects of
a cutoff in the partial wave summation for the evaluation of the
momentum distribution in finite nuclei. Comparing the prediction for
the total $\Delta n(k)$ with those employing a restriction to a maximal
orbital angular momentum, one observes that such a restriction leads to
remarkable differences in particular at high momenta. Using $L_{max} =
3$ in eq.(\ref{eq:lmax}) one obtains a prediction for the momentum
distribution, which is only one half of the result including all
partial waves at $k \approx$ 3.5 fm$^{-1}$.

The right part of fig.\ref{fig:o3} shows the comparison of the
continuum part of the momentum distribution evaluated directly for
$^{16}O$ with the LDA prediction using $L_{max} = 4$, the maximal
orbital angular momentum, which we have taken into account in our
direct evaluation for the finite system. The agreement of the LDA with
the direct evaluation is very good in particular at high momenta. For
small momenta the direct calculation for the finite system yields a
larger value than the LDA. For these small momenta, however, one must
keep in mind that the continuum part calculated directly for the finite
system contains components, which have a momentum distribution similar
to the quasihole part\cite{pap2},
whereas the $\Delta n(k)$ evaluated in LDA only accounts for
momenta above the local Fermi momenta $k_{F}(r)$.

Concluding we may say that the discrepancy observed in ref.\cite{pap2}
between the predictions of high momentum components in the
single-particle density of nuclear matter and those for finite systems
has been resolved. The LDA seems to produce very similar results if the
same NN interaction is used, an appropriate average over nuclear matter
with various densities is considered and the effects of truncating the
partial wave expansion in finite systems are taken into account.

\section{Conclusions}
The single-particle momentum distribution has been
investigated for nuclear matter and the finite nucleus employing two
different realistic models for the NN interaction. The investigations
are based on the Green's function approach approximating the
self-energy of the nucleon including all contributions up to second
order in the $G$-matrix. The main results can be summarized by the
following conclusions.
\begin{itemize}
\item The present approach yields results for nuclear matter which are
in very good agreement with the more elaborate calculations of
\cite{vonde}. This gives us some confidence that the same approximation
should also produce reliable results if applied for finite systems.
\item The momentum distribution observed in knock-out experiments with
small missing energies should mainly observe the quasihole distribution
with small components at high momenta. Larger contributions to the
single-particle density at high momenta should be obtained at large
excitation energies of the residual nucleus.
\item The prediction for the momentum distribution depend weakly but
in a rather characteristic way on the interaction used. The stronger
tensor and short-range
components of the Reid soft-core potential yields a larger
single-particle strength at high momenta and large missing energies as
compared to the OBE potential $B$ of \cite{rupr}.
For momenta around 3.5 fm$^{-1}$ the momentum distribution derived from
the Reid interaction is larger by a factor 2 in nuclear matter as well
as finite nuclei.
\item The momentum distribution calculated for finite nuclei is rather
sensitive to a truncation in a partial wave expansion. Orbits with
angular momenta $l$ larger than 4 should be taken into account to
obtain stable results at momenta $ k \approx$ 4 fm$^{-1}$.
\item A local density approximation, in which the high momentum
components in the single-particle density are derived from the study of
correlations in nuclear matter,  yields a very good agreement with
corresponding studies for finite nuclei, if the effects due to
truncations in the partial wave expansion are considered.
\item The numerical scheme developed for the solution of the Dyson
equation in nuclear matter leads to a very efficient representation of
the single-particle Green's function in terms of a few
``characteristic'' poles in the Lehmann representation. This BAGEL
approximation could be very useful e.g. in studies of nuclear response
functions beyond the HF and RPA approximations.
\end{itemize}

This research project has partially been supported by SFB 382 of the
"Deutsche Forschungsgemeinschaft", DGICYT, PB92/0761
(Spain), and the EC-contract CHRX-CT93-0323.
One of us (H.M.) is pleased
to acknowledge the warm hospitality at the Facultat de F\'\i sica,
Universitat de
Barcelona, and the support by the program for Visiting Professors of this
university.

\section{Appendix}
This appendix lists the expressions\cite{knehr} for the weighting function
$W(q)$ and the average c.m. momentum $\tilde K_{av}$ which have been used in
the calculation of the 2h1p contribution to the nucleon self-energy
according to eq.(\ref{eq:self2h1p}). For that purpose we distinguish 4
different cases of the momentum $k$ for which the self-energy shall be
evaluated:
\begin{eqnarray}
(a)&  0\leq k \leq \frac{1}{3}k_F\nonumber\\
(b)& \frac{1}{3}k_F <k < k_F \nonumber\\
(c)& k_F\leq k < 3k_F \nonumber\\
(d)& 3k_F\leq k < \infty \, .
\label{4-26}
\end{eqnarray}

\noindent
{\bf  case (a)}
\begin{equation}
\tilde {K}_{av}(q,k)=\left\{
\begin{array}{ll}
\displaystyle\frac{\frac{1}{3}\left((k+q)^3
+ (\frac{1}{2}
(k^2+k_F^2)-q^2)^\frac{3}{2}\right)}{\frac{1}{4}
(k^2-k_F^2)+q(q+k)} &
{\rm for } \; \frac{1}{2}(k_F-k)\leq q \leq \frac{1}{2}(k+k_F) \\
\rule{0em}{3em}
\displaystyle\frac{\frac{1}{3}\left(
(q+k)^3 - (q-k)^3 \right)}{ 2k q} &
{\rm for }  \;\frac{1}{2}(k+k_F) \leq q\leq (k_F-k) \\
\rule{0em}{3em}
\displaystyle\frac{\frac{1}{3}\left((k_F)^3-
(q-k)^3\right)}{\frac{1}{2}
(k_F^2-k^2-q^2)+k q} &
{\rm for } \; (k_F-k) \leq q \leq (k+k_F)
\end{array}
\right.
\label{4-31}
\end{equation}
\begin{equation}
W(q) =\left\{
\begin{array}{ll}
q\cdot\left\{\frac{1}{4}(k^2-k_F^2)+q(q+k)\right\}
& {\rm for } \frac{1}{2}(k_F-k) \leq q \leq \frac{1}{2}(k_F+k)\\

\rule{0em}{2em}q\cdot\left\{2k q\right\}
& {\rm for } \frac{1}{2}(k_F+k)\leq q \leq (k_F-k)\\

\rule{0em}{2em}q\cdot\left\{\frac{1}{2}(k_F^2-k^2-q^2)+k
q \right\}
& {\rm for } (k_F-k)\leq q \leq (k_F+k)
\end{array}\right.
\label{4-36}
\end{equation}

\noindent
{\bf case (b)}
\begin{equation}
\tilde{K}_{av}(k,q)=\left\{
\begin{array}{ll}
\displaystyle\frac{\frac{1}{3}\left((k+q)^3
+ (\frac{1}{2}
(k^2+k_F^2)-q^2)^\frac{3}{2}\right)}{\frac{1}{4}
(k^2-k_F^2)+q(q+k)} &
{\rm for }  \; \frac{1}{2}(k_F-k)\leq q \leq (k_F-k)  \\
\rule{0em}{3em}
\displaystyle\frac{\frac{1}{3}\left(
k_F^3-(\frac{1}{2}(k^2+k_F^2)-q^2)^\frac{3}{2}
\right)}{ \frac{1}{4}(k_F^2-k^2)+\frac{1}{2}q^2} &
{\rm for }  \; (k_F-k)  \leq q \leq \frac{1}{2}(k+k_F)\\
\rule{0em}{3em}
\displaystyle\frac{\frac{1}{3}\left((k_F)^3-
(q-k)^3\right)}{\frac{1}{2}
(k_F^2-k^2-q^2)+k q} &
{\rm for }  \; \frac{1}{2}(k+k_F) \leq q \leq (k+k_F)
\end{array}
\right.
\label{4-32}
\end{equation}
\begin{equation}
W(q) =\left\{
\begin{array}{ll}
q\cdot\left\{\frac{1}{4}(k^2-k_F^2)+q(q+k)\right\}
& {\rm for } \frac{1}{2}(k_F-k) \leq q \leq (k_F-k)\\

\rule{0em}{2em}k\cdot\left\{\frac{1}{4}(k_F^2-k^2)
+\frac{1}{2}q^2\right\}
& {\rm for } (k_F-k) \leq q \leq \frac{1}{2}(k_F+k)\\

\rule{0em}{2em}k\cdot\left\{\frac{1}{2}(k_F^2-k^2-q^2)+k
q \right\}
& {\rm for } \frac{1}{2}(k_F+k)\leq q \leq (k_F+k)
\end{array}\right.
\label{4-37}
\end{equation}

\noindent
{\bf case (c)}
\begin{equation}
\tilde{K}_{av}(k,q) = \left\{
\begin{array}{ll}
\displaystyle\frac{\frac{1}{3}\left(
k_F^3-(\frac{1}{2}(k^2+k_F^2)-q^2)^\frac{3}{2}
\right)}{ \frac{1}{4}(k_F^2-k^2)+\frac{1}{2}q^2} &
{\rm for } \; \sqrt{\frac{1}{2}(k-k_F)(k+k_F)}
\leq q \leq \frac{1}{2}(k+k_F)\\
\rule{0em}{3em}
\displaystyle\frac{\frac{1}{3}\left((k_F)^3-
|q-k|^3\right)}{\frac{1}{2}
(k_F^2-k^2-q^2)+k q} &
{\rm for } \; \frac{1}{2}(k+k_F) \leq q \leq (k+k_F)
\end{array}
\right.
\label{4-33}
\end{equation}
\begin{equation}
W(q)=\left\{
\begin{array}{ll}
q\cdot\left\{\frac{1}{4}(k_F^2-k^2)
+\frac{1}{2}q^2\right\}
& {\rm for }  \sqrt{\frac{1}{2}(k-k_F)(k+k_F)}
 \leq q \leq \frac{1}{2}(k_F+k)\\

\rule{0em}{2em}q\cdot\left\{\frac{1}{2}(k_F^2-k^2-q^2)+k
q \right\}
& {\rm for }  \frac{1}{2}(k_F+k)\leq q \leq (k_F+k)
\end{array}\right.
\label{4-38}
\end{equation}

\noindent
{\bf case (d)}
\begin{equation}
\tilde{K}_{av}(k,q)=\left\{
\begin{array}{ll}
\displaystyle\frac{\frac{1}{3}\left((k_F)^3-
|q-k|^3\right)}{\frac{1}{2}
(k_F^2-k^2-q^2)+k q} &
{\rm for } \; (k-k_F) \leq q \leq (k+k_F)
\end{array}
\right.
\label{4-34}
\end{equation}
\begin{equation}
W(q) = \left\{
\begin{array}{ll}
q \cdot \left\{\frac{1}{2}(k_F^2-k^2-q^2)+k
q \right\}
& {\rm for }  (k-k_F)\leq q \leq (k_F+k)
\end{array}\right.
\label{4-39}
\end{equation}

\begin{table}[h]
\caption{Distribution of nucleons in $^{16}$O.
Listed are the occupation probabilities for various
partial waves. These probabilities are obtained by integrating the
partial wave contributions in eq.(32) ignoring the degeneracy factors
$2(2j+1)$. For states with $l \leq 1$ the contributions from the
quasihole ($ n^{qh}$) and the continuum part ($n^c$) of the spectral
function are listed  separately. Results have been obtained using the
OBE potential $B$ and the Reid soft-core potential.}
\label{tab:tab1}
\begin{center}
\begin{tabular}{c|rrrr}
&&&&\\
&\multicolumn{2}{c}{OBEP $B$}&\multicolumn{2}{c}{Reid}\\
$lj$&\multicolumn{1}{c}{$ n^{qh}$}&\multicolumn{1}{c}{$ n^c$}
&\multicolumn{1}{c}{$ n^{qh}$}&\multicolumn{1}{c}{$ n^c$}\\
&&&&\\ \hline
&&&&\\
$s_{1/2}$ & 0.780 & 0.157 & 0.778 & 0.117 \\
$p_{3/2}$ & 0.914 & 0.042 & 0.896 & 0.040 \\
$p_{1/2}$ & 0.898 & 0.046 & 0.896 & 0.047 \\
$d_{5/2}$ && 0.022 & & 0.018\\
$d_{3/2}$ && 0.027 & & 0.024\\
$f_{7/2}$ & & 0.008 & & 0.014 \\
$f_{5/2}$ & & 0.013& & 0.019 \\
$g_{9/2}$ && 0.002 && 0.004 \\
$g_{7/2}$ && 0.004 && 0.007 \\
&&&&\\
\end{tabular}
\end{center}
\end{table}

\clearpage
\begin{figure}
\caption{Graphical representation of the Hartree-Fock (a), the 2-particle
1-hole (2p1h, b) and the 2-hole 1-particle contribution (2h1p, c) to the
self-energy of the nucleon}
\label{fig:diag}
\end{figure}
\begin{figure}
\caption{Momentum distribution in nuclear matter at the empirical
saturation density, $k_{F}$ =1.36 fm$^{-1}$. Results obtained with the
approximations discussed in section 2 for the OBE potential $B$
(dashed-dotted line) and the Reid soft-core potential (dashed line) are
compared to the results using the Reid potential (solid line ``exact'')
as derived from the more sophisticated calculations of ref.[18]}
\label{fig:nm1}
\end{figure}
\begin{figure}
\caption{Occupation probabilities in nuclear matter as a function of
the momentum in units of the Fermi momentum $k_{F}$. Results obtained
for the OBE potential $B$ are shown for various densities (see description
in the figure). The occupation probabilities for momenta below $k_{F}$
are displayed with reference to the axis on the left side of the
figure, while those for $k$ larger $k_{F}$ refer to the logarithmic
scale on the axis at the right side.}
\label{fig:nm2}
\end{figure}
\begin{figure}
\caption{The contribution of the quasihole state to the occupation
probability (curves labeled with triangle, see eq.(40)) and the continuum
contribution as a function of the momentum $k/k_{F}$ calculated for the
OBE potential $B$. The dashed lines refer to a Fermi momentum of
nuclear matter of 1.0 fm$^{-1}$, whereas the solid lines are obtained
for $k_{F}$ = 1.35 fm$^{-1}$.}
\label{fig:nm3}
\end{figure}
\begin{figure}
\caption{Energy spectra for nuclear matter with Fermi momentum $k_{F}$
= 1 fm$^{-1}$ (dashed curves) and $k_{F}$ = 1.35 fm$^{-1}$ (solid
lines) as a function of momentum. The left part of the figure exhibits
the HF single-particle energies (curves labeled with triangles) and
$E_{qh}$ the energies of the quasihole states. In the right part the
mean values $\bar \omega$ of the continuum part of the momentum
distribution (see eq.42) are displayed.}
\label{fig:nm4}
\end{figure}
\begin{figure}
\caption{Occupation probabilities in nuclear matter at $k_{F}$=1.35
fm$^{-1}$ using various modifications of the BHF single-particle
spectrum}
\label{fig:nm5}
\end{figure}
\begin{figure}
\caption{Occupation probabilities in nuclear matter at $k_{F}$=1.35
fm$^{-1}$ calculated in various BAGEL(n,m) approximations are compared
to the result obtained with the complete Lehmann representation of the
single-particle Green's function.}
\label{fig:nm6}
\end{figure}
\begin{figure}
\caption{Momentum distribution for $^{16}O$ in the $p_{1/2}$ partial
wave (see eq.(32)). The distributions are normalized such that
$\int dk\ n(k)\ =\ 1$ if one orbit would be occupied. The three parts
of the figure display the quasihole contribution, the continuum
contribution and the sum of these two (total) as obtained for the OBE
$B$ (solid lines) and the Reid soft-core potential (dashed lines).}
\label{fig:o1}
\end{figure}
\begin{figure}
\caption{Continuum part of the single-particle density for $^{16}O$ as
a function of the momentum (left part, see eq.(32)) and energy (right
part). Results are presented as obtained for the OBE
$B$ (solid lines) and the Reid soft-core potential (dashed lines).}
\label{fig:o2}
\end{figure}
\begin{figure}
\caption{Continuum part of momentum distribution for $^{16}O$ as
obtained from the local density approximation (see eq.(35)) using the
OBE potential $B$. In the left part of the figure the total
contribution is compared to results obtained assuming various values
$L_{max}$ for the truncation of the partial wave expansion. The right
part of the figure compares the local density approximation with
$L_{max}$=4 with the result obtained in a direct calculation of
$^{16}O$.}
\label{fig:o3}
\end{figure}

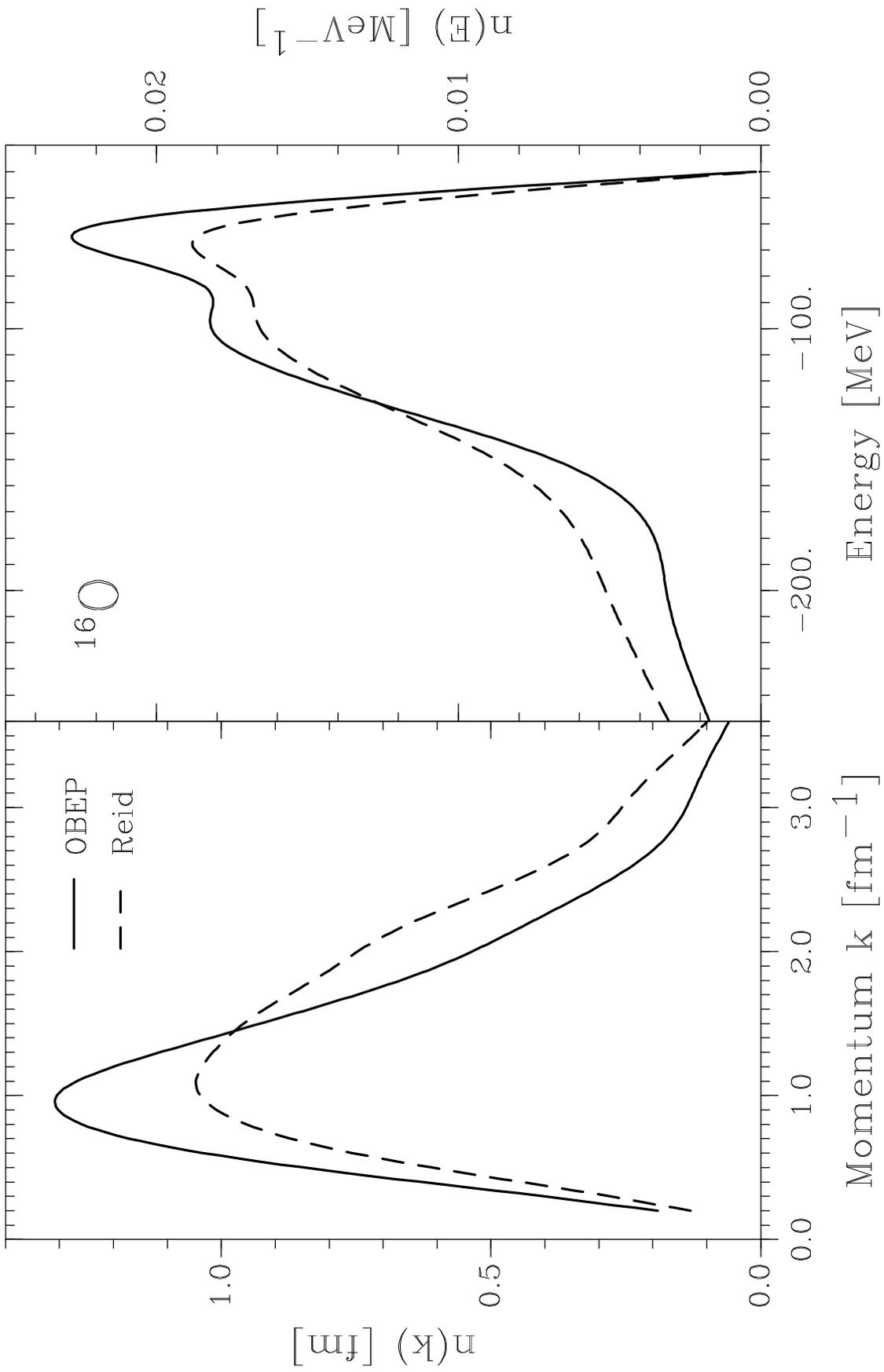
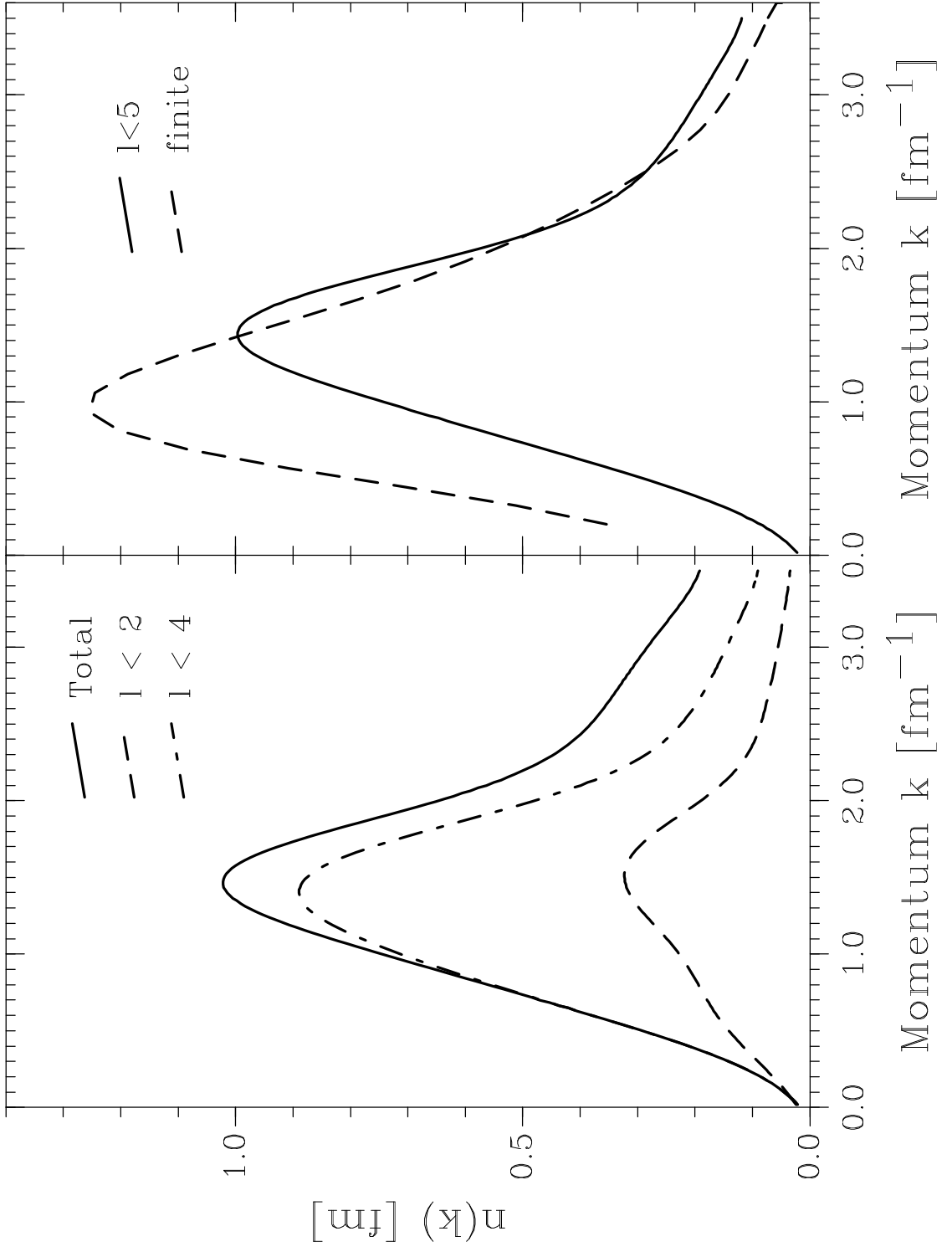
\end{document}